\documentclass{article}

{}
{}
{}

\usepackage{amsmath} 
\usepackage{amsthm} 
\usepackage{hyperref} 
\usepackage{graphics} 
\usepackage{graphicx}
\usepackage{algorithmic} 
\usepackage{hyperref}
\usepackage{authblk}
\usepackage[margin=1.2cm]{geometry}
\usepackage{multicol}
\usepackage{blindtext}
\usepackage{epstopdf}
\usepackage{caption}
\usepackage{titlesec}
\newenvironment{Figure}
  {\par\medskip\noindent\minipage{\linewidth}}
  {\endminipage\par\medskip}
\usepackage{lipsum}
\usepackage{titlesec}
\usepackage{biblatex}
\titlespacing*{\section}{0pt}{0.1\baselineskip}{0.2\baselineskip}
\begin{document}

\title{\textbf{Redactable and Sanitizable Signature Schemes: Applications and Limitations for use in Decentralized Digital Identity Systems}}

\author{Bryan Kumara}
\author{Mark Hooper}
\author{Carsten Maple}
\author{Timothy Hobson}
\author{Jon Crowcroft}
\affil{{The Alan Turing Institute, London, U.K.} Email: bkumara, mhooper, cmaple, thobson, jcrowcroft @turing.ac.uk}
   
\date{18 April 2023}

\renewcommand\Affilfont{\itshape\small}

\maketitle

\begin{abstract}
Redactable signature schemes and sanitizable signature schemes are methods that permit modification of a given digital message and retain a valid signature. This can be applied to decentralized identity systems for delegating identity issuance and redacting sensitive information for privacy-preserving verification of identity. We propose implementing these protocols on a digital credential and compare them against other privacy-enhancing techniques to assess their suitability.
\end{abstract}

\begin{multicols}{2}

\section{Introduction}
  
A critical aspect of economic growth is the integration of the internet as a medium for innovation \cite{Ali:2013}. It is pivotal that that digital identities, which refer to information used by computers to represent external agents, address stakeholder concerns \cite{Ali:2013} \cite{Camp:2004} \cite{Atick:2016}. Every digital identity system comes with its own set of trade-offs that considers security, reliability, resilience, accessibility, and scalability amongst many other factors \cite{digital}.
\\ A key issue is the degree on which a single party may exert control over the operation of the system. Centralized digital identity systems usually contain a dedicated registration process, unique user identifiers, and the storage of information in a central database \cite{Fang:2009}. However, the centralization of authority and storage carries the risk of abuse such as exclusion and censorship alongside individual profiling \cite{Ellemers:2002}. 
\\ In contrast, decentralized identity systems typically approach the issue using transparent, open networks and protocols \cite{Dib:2020}. They usually involve voluntary participation and attempt to minimize data collection or formal enrolment. In this approach, standards such as Decentralized Identifiers and Verifiable Credentials help prove set membership to establish that individuals possess certain attributes without needing a central authority \cite{Lux:2020}.
\\ This extended abstract examines how Redactable and Sanitizable Signature Schemes can enable privacy preserving computation in the context of decentralized identity systems and consider mitigating solutions to some of their limitations. Redactable and Sanitizable Signature Scheme can be applied to anonymous disclosure and delegating identity issuance respectively. We propose implementing these techniques for a decentralized identity system to better understand their performance against other privacy-enhancing techniques.

\section{Background}
Digital signature allow users to verify the authenticity of a message \cite{kaur:2012}. A valid signature for a given message proves the message was sent by a known user and has not been altered during transit \cite{kaur:2012}. Traditional digital signatures become invalid if even a single part of the associated message is modified \cite{kaur:2012}. However, in some cases users may want to allow a third party to alter a signed message and still retain a valid signature such as removing personal details of patient data for a medical study \cite{bilzhause:2017}. To support this, Sanitizable Signature Schemes (SSS) allow a designated third party, known as the sanitizer, to change permissible parts of a signed message. Alternatively, Redactable Signature Schemes (RSS) allow parties to redact parts of a signed message \cite{bilzhause:2017}. RSS and SSS are not recent, however research into these constructions are relatively limited compared to other privacy-enhancing techniques such as multiparty computation and zero-knowledge proofs. For the multitude constructions of RSS and SSS, readers should refer to \cite{bilzhause:2017}. In general, these schemes are based on pairing-based cryptography which maps elements between pairs of cryptographic groups. RSS and SSS gather the altered parts of a message and commit them cryptographically, generating a derived signature from the original signature. A verification process would confirm that this derived signature is well-formed.
\\Both RSS and SSS share standard security properties including being unforgeable, the prevention of deriving information of sanitized or redacted parts, the immutability of inadmissible parts of a message, and accountability of who modifies a message. Various constructions offering different properties can be found in \cite{bilzhause:2017}.

\section{Applications to Decentralized Digital Identity}
Though RSS and SSS have a broad range of applications, they can be particularly useful in decentralized identity systems. The former can be applied for Verifiable Credentials (VC) which is an open standard for digital credentials that can represent information such as bank account ownership and licenses by enabling selective disclosure \cite{Sedlmeir:2021}. For instance, the issuer of a bank account ownership would be a bank, and this information can be held by the bank account owner who becomes the holder of a credential. An associated digital signature attests to this issuance. The holder manages the credentials and can present it to verifiers such as mortgage companies as needed. Verifiers can request the VC from the holders as seen in the figure below \cite{Kang:2022}. A key step is to check if the signature of the issuer is valid, which indicates message authenticity and integrity. However, holders may not want to disclose the entire VC content on every request, as it may contain sensitive information depending on the format of the digital identity.

\begin{Figure}
 \centering
 \includegraphics[width=\linewidth]{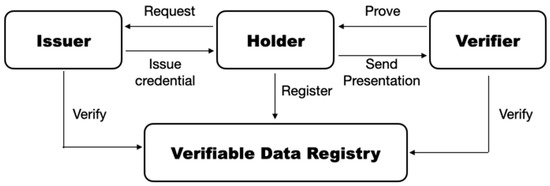}
 \captionof{figure}{DID Model proposed by W3C \cite{Kang:2022}}
\end{Figure}

One option to limit VC disclosure would be using zero-knowledge proofs to allow a party to show that a given statement is true, such as being in possession of a certain amount of cash, while  not revealing any additional information beyond the fact that the said condition is met \cite{Maurer:2009}. However, zero-knowledge proofs are complicated techniques with multiple variants. Some require trusted setups which may not be suitable for decentralized identity systems with various representations and proof techniques \cite{Sun:2021}. RSS allows a holder to redact segments of a VC but leave other components unredacted for verification. Current theoretical performance speed of RSS indicates that this is the more efficient solution due to their speed and smaller signature size, while proof of knowledge protocols may have sizes larger than the signature of an unredacted VC.
\\ From a credential issuer’s perspective, decentralization may require some form of delegation for scalability. Decentralized Identifiers (DIDs) are a type of unique identifier that is persistent, verifiable, and without the need of a centralized registry which enables a new model of decentralized digital identity. The World Wide Web Consortium (W3C), the main international standards organization for the World Wide Web, maintains an open standard for DIDs. Often, DID issuers will want to delegate authority by allowing other nodes to issue DIDs with the delegator’s approval. However, to avoid malicious behaviour from delegates, such generating dud DIDs which may lead to Sybil attacks, there needs to be accountability. One approach is to use a multi-signature scheme which requires the signature of multiple parties to jointly sign a proposed DID \cite{Nick:2021}. Using this method the delegate node would gather the necessary information but cannot immediately issue a DID as it requires a superior’s signature before being issued. 
\\ However, one weakness of this approach is the back-and-forth communication required between a delegate and the delegating node. Every proposed DID needs to be checked by the delegating node that it has been formatted correctly before obtaining their signature to create the DID. In areas of low connectivity and or down-time of either party, there will be a delay in issuing DIDs from a delegate node. SSS allows a delegating node to give a delegate blank templates that can be modified and some SSS allow sanitization in a constrained fashion, this will ensure that a potential DID will follow the correct format. If the downstream entity formats the DID correctly, a valid signature can be automatically derived for the DID without another round of communication. Like RSS, the performance time of SSS is reasonably quick and is helped by the relatively small number fields in a W3C DID standards. Of course, SSS can also be applied to other hierarchical models where delegation of authority is needed, such as in Public Key Infrastructure, where root certificate authorities delegate authorities to subordinate certificate authorities. 

\section{Limitations and Mitigations}
SSS and RSS possess limitations that need to be addressed. Using SSS, a malicious delegate may generate multiple dud instances that are correctly filled. Neither SSS nor multi-signatures address the veracity of the input, and this issue of a single-source oracle for decentralized system is a problem outside the scope of this abstract. However, the potential to generate multiple dud instances is somewhat limited in a multi-signature scheme. If the issuer receives an unusually high number of requests to approve DIDs, it may indicate malicious activity and could refuse the multi-signature request. SSS does not have traffic monitoring as sanitization requires no communication between the two parties outside of the original issuance of sanitizable message. One solution is to use threshold cryptography, where each DID a delegate creates reveals a share of their secret key. Crossing the threshold by generating too many DIDs remove the autonomy of the delegate. Though this approach seems promising, depending on the threshold it can incur a high computational cost \cite{Canard:2010}.

Another general issue of the two is the lack of post-quantum research into the field. Whilst other techniques such as zero-knowledge proofs and multi-party computations have multiple instances and analysis into post-quantum variants, there is a distinct lack of them for RSS and SSS. Only very recently, RSS has had a lattice-based \cite{Zhao:2022} and a hash-based construction \cite{Zhu:2022} proposed; more research will be needed to understand security and performance implications. SSS has no post-quantum variant that the author is aware of, however the use of the post-quantum SPHINCS+ scheme in RSS \cite{Zhu:2022} is promising and may be used in constructions of SSS that uses chameleon hashes to permit sanitization. 
\\ Further work is required to examine SSS and RSS benefits for decentralized identity systems. We aim to use RSS and SSS on DIDs based on the W3C standard to examine the size and speed of these schemes compared to alternatives such as zero-knowledge proofs and multi-signatures.

\section{Conclusion}
Decentralized identity systems can implement selective disclosure and delegate identity issuance using RSS and SSS respectively. These related schemes permit the modification of a message whilst retaining a valid signature, providing an alternative to techniques such as zero-knowledge proofs and threshold cryptography. However, a lack of research into the two needs to be remedied before widespread adoption is possible. We propose future work examining the performance of these schemes against other privacy-enhancing techniques.

\section{Acknowledgments}
This work was supported, in whole or in part, by the Bill and Melinda Gates Foundation [INV-001309]. Under the grant conditions of the Foundation, a Creative Commons Attribution 4.0 Generic License has already been assigned to the Author Accepted Manuscript version that might arise from this submission.

\printbibliography 

\end{multicols}
\end{document}